**Persistent semi-metal-like nature of epitaxial perovskite CaIrO$_3$ thin films**

Abhijit Biswas, and Yoon Hee Jeong[*]

Department of Physics, POSTECH, Pohang, 790-784, South Korea

**Abstract**

Strong spin-orbit coupled 5$d$ transition metal based $AB$O$_3$ oxides, especially iridates, allow tuning parameters in the phase diagram and may demonstrate important functionalities, for example by means of strain effects and symmetry-breaking, because of the interplay between the Coulomb interactions and strong spin-orbit coupling. Here, we have epitaxially stabilized high quality thin films of perovskite (Pv) CaIrO$_3$. Film on the best lattice-matched substrate shows semi-metal-like characteristics. Intriguingly, imposing tensile or compressive strain on the film by altering the underlying lattice-mismatched substrates, still maintains semi-metallicity with minute modification of the effective correlation as tensile (compressive) strain results in tiny increases (decreases) of the electronic bandwidth. In addition, magnetoresistance remains positive with a quadratic field dependence. This persistent semi-metal-like nature of Pv-CaIrO$_3$ thin films with minute changes in the effective correlation by strain may provide new wisdom into strong spin-orbit coupled 5$d$ based oxide physics.

[*]Corresponding author: *yhj@postech.ac.kr*

# I. INTRODUCTION

Since the discovery of an emergent $J_{eff} = 1/2$ Mott insulating ground state in Sr$_2$IrO$_4$ [1], strong spin-orbit coupled 5$d$ transition metal based oxides have become one of the hottest topics in the field of condensed-matter physics showing a plethora of unique physical properties such as dimensionality controlled metal-insulator-transitions, unconventional magnetic ordering, spin-liquid behavior, and topological insulator characteristics.[2] Over the past few years, Ruddlesden-Popper phases of strontium iridates, Sr$_{n+1}$Ir$_n$O$_{3n+1}$ ($n$ = 1, 2, …, ∞) have been investigated with widespread enthusiasm, as increasing $n$, i.e. increasing the number of IrO$_2$ planes, changes its $J_{eff} = 1/2$ and $J_{eff} = 3/2$ band picture with the systematic transition from insulator-to-metal, because of the interplay between local Coulomb interactions, strong spin-orbit coupling, and dimensionality.[1,3,4]

Recently, the analogous compound CaIrO$_3$, has attracted a lot attention due to its geophysical significance as its low dimensional *post-perovskite* (pPv) structure having space group-*Cmcm*, was found to be similar with the archetype phase of the pPv polymorph of MgSiO$_3$, found in Earth's D" layer.[5,6] The pPv-CaIrO$_3$ is naively expected to be a good metal having 5$d$ element as 5$d$ orbitals are spatially more extended than its 3$d$,4$d$ counterpart giving larger bandwidth [3]. However, it actually exhibits a Mott insulating $J_{eff} = 1/2$ state with resistivity ($\rho$) ~ 10$^4$ Ω·cm at $T$ = 300 K.[7] Furthermore, it undergoes a canted antiferromagnetic phase transition at $T$ = 115 K.[7] It has been concluded that strong spin-orbit coupling as well as Coulomb interactions play a crucial role in forming the $J_{eff} = 1/2$ insulating state as well as in the observed magnetism.[8-10] In contrast, three dimensional *perovskite* (Pv) CaIrO$_3$ having space group-*Pbnm*, shows a large drop ($\sim 10^5$) in resistivity with respect to its pPv form, nevertheless, resistivity increases with the decrease in temperature.[7] Surprisingly, the resistivity does not diverge even at low temperature showing semi-metal-like nature, and the system also shows Pauli paramagnetic characteristics.[7] However, although polycrystalline Pv-CaIrO$_3$ was shown to be semi-metal-like and paramagnetic, from materials science perspective, in view of the progress in oxide thin film heterostructures fabrication by epitaxial methods, takes advantages over bulk methods because crucial parameters such as bandwidth (*W*), correlation *(U)* and the strength of spin-orbit coupling can be changed to some degree by tailoring substrate-induced epitaxial strain, utilizing structurally compatible different lattice-mismatched substrates and thus Pv-CaIrO$_3$ thin films would also be able to provide new interesting phenomena, as similar to that in Pv-SrIrO$_3$ thin films.[11-13]

In recent years, few efforts were given to synthesize Pv-CaIrO$_3$ thin films as the depositions of iridates in general are very difficult, because the stoichiometry is difficult to maintain since the Pv structure is not the thermodynamic stable phase and can be stabilized over the thermodynamically favored polymorph with edge-shared octahedra only at high pressure and temperature.[14] Despite these limitations, however, by taking advantage of thin film growth technology Pv-CaIrO$_3$ films were grown and found to be semi-metal-like.[15,16] As a matter-of-fact, the system lies close to the metal-insulator-transition phase boundary. This aspect is particularly significant as the metal-insulator-transition and many other transport properties in a correlated system are controlled by competition among relevant energy scales indicating *W, U,* and spin-orbit coupling.[3,15] The key parameter governing the properties of epitaxial films is mainly the effective correlation (*U/W*), caused by a change in electronic bandwidth (*W*) due to octahedral distortion originated from the lattice-mismatch between the film and the substrate. Consequently, it can be anticipated that applying *strain* by growing Pv-CaIrO$_3$ thin films on different lattice-mismatched substrates would bring about corresponding changes in the octahedral distortion and thus its physical and electrical properties, especially the metal-insulator-transition scenario.

The electronic bandwidth (*W*) of a Pv-ABO$_3$ system changes as $W \propto \frac{\cos \varphi}{d^{3.5}}$, where *d* is the *B-O* bond length and $\varphi = (\pi - \theta)/2$ is the buckling deviation of the *B-O-B* bond angle *θ* from π.[17] In principle, on a hetero-epitaxial thin film, an externally imposed strain by using different lattice-mismatched substrates would cause a change in the electronic bandwidth as the *B-O-B* bond angle as well as *B-O* bond length changes. It was often found that for Pv thin films, the in-plane strain is mostly adopted by tilting of corner-connected BO$_6$ octahedra or a change in bond angle.[18,19] However, in general, the relationship between the strain in the film and its electronic bandwidth change may not be so direct if we consider the biaxial nature of the strain in the film.[20-22] Nevertheless, the average change in bond length and bond angle would be sufficient to result in corresponding change in the electronic bandwidth (*W*) and therefore the electronic transport properties. The understanding of which had been one of the main motivations started the field of condensed-matter physics.

In this report, we grew epitaxial Pv-CaIrO$_3$ thin films and made a systematic study of their electrical properties. Starting with the film on best lattice-matched substrate (most natural state), we then altered the underlying substrates as these imposed high tensile or compressive strain on the film. Intriguingly, the semi-metal-like nature in resistivity persists irrespective of the imposed strain (tensile/compressive) with minute changes in resistivity. Persistency in semi-metal-like characteristics implies that the effective correlation ($U/W$) is thus minutely modified by substrate-induced epitaxial strain on Pv-CaIrO$_3$ thin films.

## II. EXPERIMENTAL DETAILS

We grew Pv-CaIrO$_3$ thin films by pulsed laser deposition (KrF excimer laser with a wavelength of 248 nm); ∼40 nm thin films were grown on various lattice-mismatched substrates (YAlO$_3$ (110), LaAlO$_3$ (001), NdGaO$_3$ (110), SrTiO$_3$ (001), and GdScO$_3$ (110)). Pv-CaIrO$_3$ thin films were fabricated from a polycrystalline target made by mixing the stoichiometric amount of raw powders (CaCO$_3$ and IrO$_2$), pelletizing, and sintering at 1000° C and ambient pressure. For growth the laser was operated at a frequency of 4 Hz, and the substrate temperature and oxygen partial pressure were 600 °C and 20 mTorr, respectively. All the films were post annealed at the same growth temperature and oxygen partial pressure for 30 min to compensate for any oxygen deficiency. To check the crystalline quality, X-ray diffraction (XRD) measurements were performed by the Empyrean XRD System from PANalytical. Atomic force microscopy (AFM) was used to determining surface topography. These studies were done with XE-100 Advanced Scanning Probe Microscope. Elemental distributions were evaluated by secondary ions mass spectrometry (SIMS) with a primary beam source of O$_2^+$ with impact energy of 7.5 keV. Electrical transport measurements were performed using the four-probe van der Pauw geometry.

## III. RESULTS AND DISCUSSION

Bulk Pv-CaIrO$_3$ is of orthorhombic structure ($a$ = 5.35046 Å, $b$ = 5.59291 Å, $c$ = 7.67694 Å)**[23]**, and orthorhombic indices were typically used to indicate the orientation of substrates as in YAlO$_3$ (110), NdGaO$_3$ (110) and GdScO$_3$ (110). However, it was more convenient for mutual comparison to use *pseudo-cubic* and cubic lattice parameters with subscript *'c'*. The *pseudo-cubic* ($a_c$) lattice parameter was converted from the orthorhombic lattice parameters ($\sqrt{2} \times \sqrt{2} \times 2$) of bulk Pv-CaIrO$_3$ and we calculated $a_c \sim$ 3.86 Å.**[23,24]** If one takes this value literally, it matches the NdGaO$_3$ substrate ($a_c \sim$ 3.86 Å) and corresponds to +3.76%, and +1.85% compressive strain for the films grown on the YAlO$_3$ ($a_c \sim$ 3.72 Å), and LaAlO$_3$ substrates ($a_c \sim$ 3.78 Å); -1.15%, and -2.52% tensile strain for the films grown on the SrTiO$_3$ ($a_c \sim$ 3.905 Å), and GdScO$_3$ substrates ($a_c \sim$ 3.96 Å), respectively. The in-plane lattice parameter of the substrate was in relationship to that of Pv-CaIrO$_3$ (Fig. 1). The substrates range conveniently from best lattice-matched (NdGaO$_3$) to highly lattice-mismatched (GdScO$_3$ and YAlO$_3$).

X-ray diffraction (XRD) measurements ($\theta$-$2\theta$ scan) of the Pv-CaIrO$_3$ films grown on the aforementioned substrates show a crystalline (001)$_C$ peak without any impurity or additional peaks; for clarity, only the low angle data were shown (Fig. 2a). Films on all the substrates (except those grown on LaAlO$_3$) exhibit clear layer thickness fringes (results of coherent scattering from a finite number of lattice planes with a coherent film thickness) confirming the homogeneous nature of the film surface as well as high crystalline quality. In contrast, the dense twinning of the LaAlO$_3$ substrate was probably responsible for degrading the surface quality of films grown on LaAlO$_3$ and the disappearance of layer thickness fringes. To further characterize the structural quality, we looked at the perfect lattice-matched film (grown on NdGaO$_3$) and measured the reflectivity (which showed nice oscillations; Fig. 2b) as well as the rocking curve ($\omega$ scan; Fig. 2c). A small full width at half maximum (FWHM $\sim$ 0.032°) from the rocking curve analysis around the (001)$_C$ peak of Pv-CaIrO$_3$ film, indicates good crystalline quality. Secondary ion mass spectrometry (SIMS) was employed to check the homogeneity in the distribution of cationic elements within the film (grown on NdGaO$_3$). The elemental distributions (Ca, Ir, and O) were homogeneous (Fig. 2d) at the resolution level of SIMS measurements. Atomic force microscopy (AFM) showed a flat surface with roughness ~0.8 nm (inset: Fig. 2d). Furthermore, we measured the strain in the Pv-CaIrO$_3$ films grown especially on the GdScO$_3$ and YAlO$_3$ substrates (highest lattice-mismatch) utilizing reciprocal space mapping (RSM). The RSM results around the in-plane (103)$_C$ reflection peak of the Pv-CaIrO$_3$ films are shown in Fig. 2e. Film grown on the GdScO$_3$ substrate (-2.52% tensile strain) was partially relaxed [15], whereas film grown on the YAlO$_3$ substrate (+3.76% compressive strain) was relaxed. It's a common feature for thin films because highly strained epitaxial films try to relax to minimize the accumulating strain energy, and this increases with the lattice-mismatch.

The measured electrical resistivity of epitaxial Pv-CaIrO$_3$ films grown on various lattice-mismatched substrates is shown within the temperature range of 10 K ≤ $T$ ≤ 300 K in Fig. 3. Film on the best lattice-matched substrate (grown on NdGaO$_3$) shows $\rho$ ~ 4 mΩ·cm at $T$ = 300 K, which is much smaller than the bulk polycrystalline Pv-CaIrO$_3$ ($\rho$ ~ 100 mΩ·cm at $T$ = 300 K) [7]; the measured resistivity in bulk polycrystalline Pv-CaIrO$_3$ was affected by large grain boundaries and porosity. The resistivity showed a tiny increase with the decrease in temperature down to 10 K. Nevertheless, $\rho(T)$ did not diverge even at low $T$ (also the change in resistivity $\frac{\rho_{300K}}{\rho_{10K}}$ was negligible, ~0.8). This is a signature of semi-metal-like characteristics.[15] This nondivergent resistivity also indicates that the band gap is not fully open and Fermi level may be located inside a pseudogap of the lower Hubbard band (*LHB*) and the upper Hubbard band (*UHB*) of $J_{eff} = 1/2$. Furthermore, Hall effect measurements confirmed the semi-metal-like behavior as the carrier concentration ($n$) was ~ 10$^{20}$ cm$^{-3}$ at $T = 300$ K; characteristic of a semi-metal. Increasing the *tensile* strain on the film (grown on SrTiO$_3$), reduces the room temperature resistivity ($\rho \sim 3.3$ mΩ·cm at $T = 300$ K) with respect to the best lattice-matched film (grown on NdGaO$_3$, $\rho \sim 4$ mΩ·cm at $T = 300$ K). Imposing more tensile strain on the film (grown on GdScO$_3$) further reduces the resistivity, although $\rho$ reached only $\sim 3$ mΩ·cm at $T = 300$ K. The overall resistivity features of these two films were found to be almost similar to the film on the best lattice-matched substrate within the temperature range of 10 K $\leq T \leq$ 300 K; these films were also characterized as semi-metal-like. In contrast, imposing *compressive* strain, by growing films (on LaAlO$_3$ and then a more compressive film on YAlO$_3$) shows an increase in resistivity with respect to the best lattice-matched film (grown on NdGaO$_3$) as $\rho$ reaches $\sim 6$ mΩ·cm at $T = 300$ K. Nevertheless, the films still remained semi-metal-like as $\rho(T)$ did not diverge at low $T$. These observations definitely have implications. However small the effect was, the effective correlation ($U/W$) modified the bandwidth of the lower Hubbard band (*LHB*) and the upper Hubbard band (*UHB*) of $J_{eff} = 1/2$ and thus moderate shift in the position of Fermi level, as strain had an impact on the *Ir-O-Ir* bond angle and *Ir-O* bond length and thus caused a change in the electronic bandwidth (*W*).**[18]** Although reasonable, there is another possible scenario; increased resistivity in a compressive strained film may also be due to disorder (structural defects, stoichiometry variations, inter-diffusion between film and the substrate), as the lattice-mismatch induced strain energy was fully released at the interface between the Pv-CaIrO$_3$ film and the substrate (YAlO$_3$) as the film was fully relaxed (Fig. 2e). However, this can be contradicted by noting that: (1) the compressively strained film was still of epitaxial nature (visible thickness fringes), and saliently (2) in case of tensile strained films, we observed a systematic decrease in resistivity, an opposed to the effect of a disorder related enhancement in resistivity. Addressing this issue is critical for further exploration of the "semi-metal-like" physics of Pv-CaIrO$_3$ thin films. In the future, perhaps, it would be desirable to observe the angle-resolved photoemission spectroscopy (ARPES) to verify the electronic structure of the Pv-CaIrO$_3$ thin film, as well as the phase diagram of Pv-CaIrO$_3$ by tuning the correlation (strain) and spin-orbit coupling; as similar done for Pv-SrIrO$_3$.[4,25]

In principle, since the iridates properties are sensitive to effective correlation, it is instructive to compare the electronic phases (resistivity wise) in Pv-CaIrO$_3$ and Pv-SrIrO$_3$ as the *Ir–O–Ir* bond angle and *Ir–O* bond length are different due to a different ionic radius on A-site. To examine this empirically, we compared the measured electrical resistivity of epitaxial Pv-SrIrO$_3$ and Pv-CaIrO$_3$ thin films grown on the respective best lattice-matched substrate (most natural state) (Fig. 4). Pv-SrIrO$_3$ film (grown on GdScO$_3$) remained fully metallic (a decrease in resistivity with the decrease in temperature) in whole temperature range (10 K $\leq T \leq$ 300 K). In contrast, the resistivity of Pv-CaIrO$_3$ thin film (grown on NdGaO$_3$) increases with the decrease in temperature but $\rho$ (T) did not diverge even at low *T*. Comparing the values, at *T* = 300 K, Pv-SrIrO$_3$ showed $\rho$ ~1.5 m$\Omega$·cm whereas for Pv-CaIrO$_3$, $\rho$ was ~4 m$\Omega$·cm. Also the resistivity of Pv-CaIrO$_3$ was higher than (~5 times) that of Pv-SrIrO$_3$ at *T* = 10 K. Within this picture, it stands to reason that the electronic bandwidth is affected by the change in effective correlation (*U/W*), as the ionic radii is smaller for Ca$^{2+}$ than Sr$^{2+}$ (Ca$^{2+}$ ~1.00 Å and Sr$^{2+}$ ~1.18 Å).[26] The smaller cationic radius of Ca$^{2+}$ would then reduce the *Ir-O-Ir* bond angle and *Ir-O* bond length in Pv-CaIrO$_3$ with respect to that of Pv-SrIrO$_3$ and hence increase the octahedral rotation in Pv-CaIrO$_3$. In fact, it has been found that in bulk polycrystalline Pv-*A*IrO$_3$ (*A* = Ca, Sr), the average *Ir-O-Ir* bond angle decreases from 154.1° (Pv-SrIrO$_3$) to 145.5° (Pv-CaIrO$_3$) while keeping the average *Ir-O* bond length nearly constant.[27] The effective correlation (*U/W*) was then enhanced for Pv-CaIrO$_3$ with respect to that of Pv-SrIrO$_3$ as the electronic bandwidth (*W*) reduces and therefore increases in resistivity.

The magnetoresistance (MR), defined as $\frac{\rho(B)-\rho(0)}{\rho(0)}$, i.e. change in electric resistivity under the influence of an external magnetic field (*B*) is an important observable phenomenon. With its many intrinsic physical aspects, sign of the MR often gives an indication of the magnetic ordering; a magnetically ordered material shows a negative MR where applying magnetic field, electron spins become ordered and thus reduce scattering.**[28]** A disordered system also shows a negative MR due to localization phenomena at low *T*.**[29]** To investigate the spin fluctuations and/or disorder effect, if any, in epitaxial Pv-CaIrO$_3$ thin films, we measured the MR of films (grown on GdScO$_3$ and on YAlO$_3$; highest lattice-mismatch), with an applied magnetic field up to $\pm 9$ T along the out-of-plane configuration at *T* = 10 K. In both cases, we observed a positive MR proportional to $B^2$ (Fig. 5). This quadratic field dependence in the MR in both cases may be attributed to the Lorentz contribution associated with the orbital motion of carriers. Most importantly, we did not observe any unusual features related with spin fluctuations or disorder in the MR up to $\pm 9$ T. A positive MR also indicated that at low *T* unusual behaviors in resistivity (shoulders at low *T*; Fig. 3) were not associated with any kind of long-range magnetic ordering; rather they were more consistent with semi-metal-like picture. This certainty had further been confirmed with magnetization measurements (not shown here) which also showed no long-range ordering within 10 K $\leq T \leq$ 300 K (as similar in polycrystalline Pv-CaIrO$_3$ **[7]**). Since the magnetic moment was very low (as expected for a system having a weak correlation and strong spin-orbit interaction), more conclusive evidence about magnetism in Pv-CaIrO$_3$ thin film needs further investigation at the microscopic level.

**IV. CONCLUSIONS**

To conclude, we have grown epitaxial Pv-CaIrO$_3$ thin films on various lattice-mismatched substrates and made a systematic study of the electrical properties. Film on the best lattice-matched substrate (most natural state) shows semi-metal-like characteristics. This semi-metal-like behavior persists even by applying large tensile or compressive strain, thus making minute modifications to the effective correlation. In addition, the films showed positive quadratic field dependent MR, without any spin fluctuations or long-range ordering. This persistent semi-metal-like nature with tiny modification of the effective correlation by strain may provide emerging physical insights as the symmetry-protected Dirac nodes in the $J_{eff} = 1/2$ bands near the Fermi level was anticipated for producing the semi-metallic state in a 5$d$ based perovskites, because of the inherent interplay of the lattice structure and large spin-orbit coupling.[25] Also in topological context, persistent semi-metal-like nature irrespective of the applied compressive or tensile strain, might be significant as the rotation and tilting angles of the oxygen octahedra determines the size of the nodal ring (Fermi surface).[25,30] Although, the above qualitative explanations may not be conclusive enough based only on transport analysis, they demands microscopic experimental observations (ARPES and/or other spectroscopy) as well as theoretical calculations (LDA+$U$+SOC). Our results will pave the way to understand this rapidly developing, yet poorly understood field of strong spin-orbit coupled 5$d$ based oxide physics.

## ACKNOWLEDGMENTS

We would like to thank Y. -W. Lee and S. -W. Kim for technical helps. Authors would also like to thank KBSI Daegu for RSM measurements. This work was supported by the National Research Foundation via SRC at POSTECH (2011-0030786).

**Figure Captions**

**FIG. 1.** (Color Online) *Pseudo-cubic* (or cubic) lattice constants of perovskite (Pv) $CaIrO_3$ and used substrates: $YAlO_3$ (110), $LaAlO_3$ (001), $NdGaO_3$ (110), $SrTiO_3$ (001) and $GdScO_3$ (110). Corresponding amount of strain is also shown; tensile (–ve) and compressive (+ve).

**FIG. 2.** (Color Online) X-ray $\theta$-$2\theta$ scan of epitaxial Pv-$CaIrO_3$ thin films grown on different substrates. Only low angle *pseudo-cubic* $(001)_C$ peaks were shown for clarity. Thickness fringes were clearly visible indicating high crystalline quality. (b)-(c) Reflectivity and full width at half maximum (FWHM) of film grown on the best lattice-matched substrate (on $NdGaO_3$) were shown. (d) Secondary ion mass spectrometry (SIMS) showed homogeneous distribution of Ca, Ir and O within the film (grown on $NdGaO_3$). Surface roughness was ~ 0.8 nm, as revealed by atomic force microscopy (inset) (e) Reciprocal space mapping (RSM) of films on high lattice-mismatched substrates are also shown. The film grown on the $GdScO_3$ (110) substrate was almost fully strained but relaxed nature was observed for film grown on $YAlO_3$ (110) substrate.

**FIG. 3.** (Color Online) Temperature dependence of the electrical resistivity $(\rho)$ of epitaxial

Pv-CaIrO$_3$ films grown on different substrates. In all cases $\rho(T)$ did not diverge even at low $T$, indicating the semi-metal-like characteristics of Pv-CaIrO$_3$ films.

**FIG. 4.** (Color Online) Comparison of electrical resistivity of epitaxial Pv-CaIrO$_3$ and Pv-SrIrO$_3$ thin films grown on the respective best lattice-matched substrate (CaIrO$_3$, NdGaO$_3$: $a_c$ ~ 3.86 Å and SrIrO$_3$, GdScO$_3$: $a_c$ ~ 3.96 Å).

**FIG. 5.** (Color Online) Positive magnetoresistance (MR) $\propto B^2$ measured for Pv-CaIrO$_3$ films (grown on GdScO$_3$ and on YAlO$_3$) at $T = 10$ K, with an applied magnetic field strength upto $\pm 9$ T, along the out-of-plane configuration.

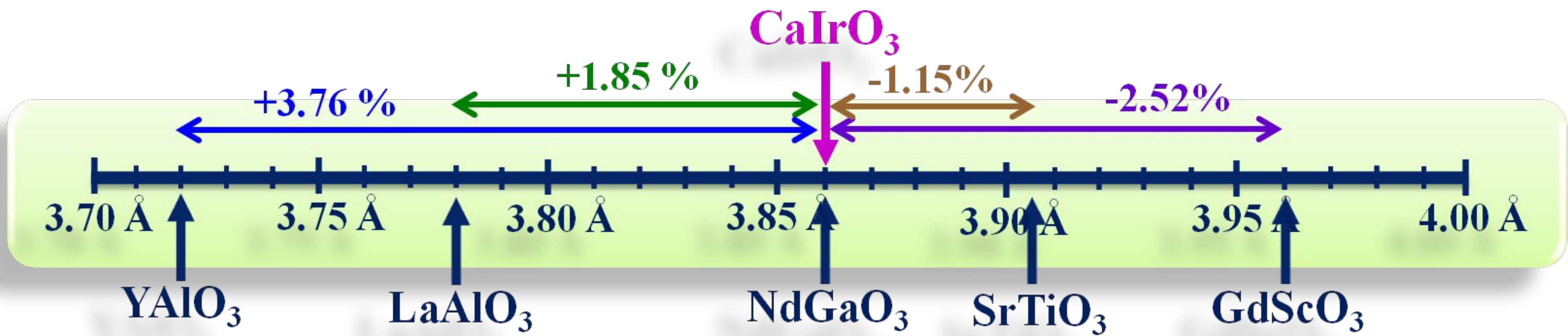

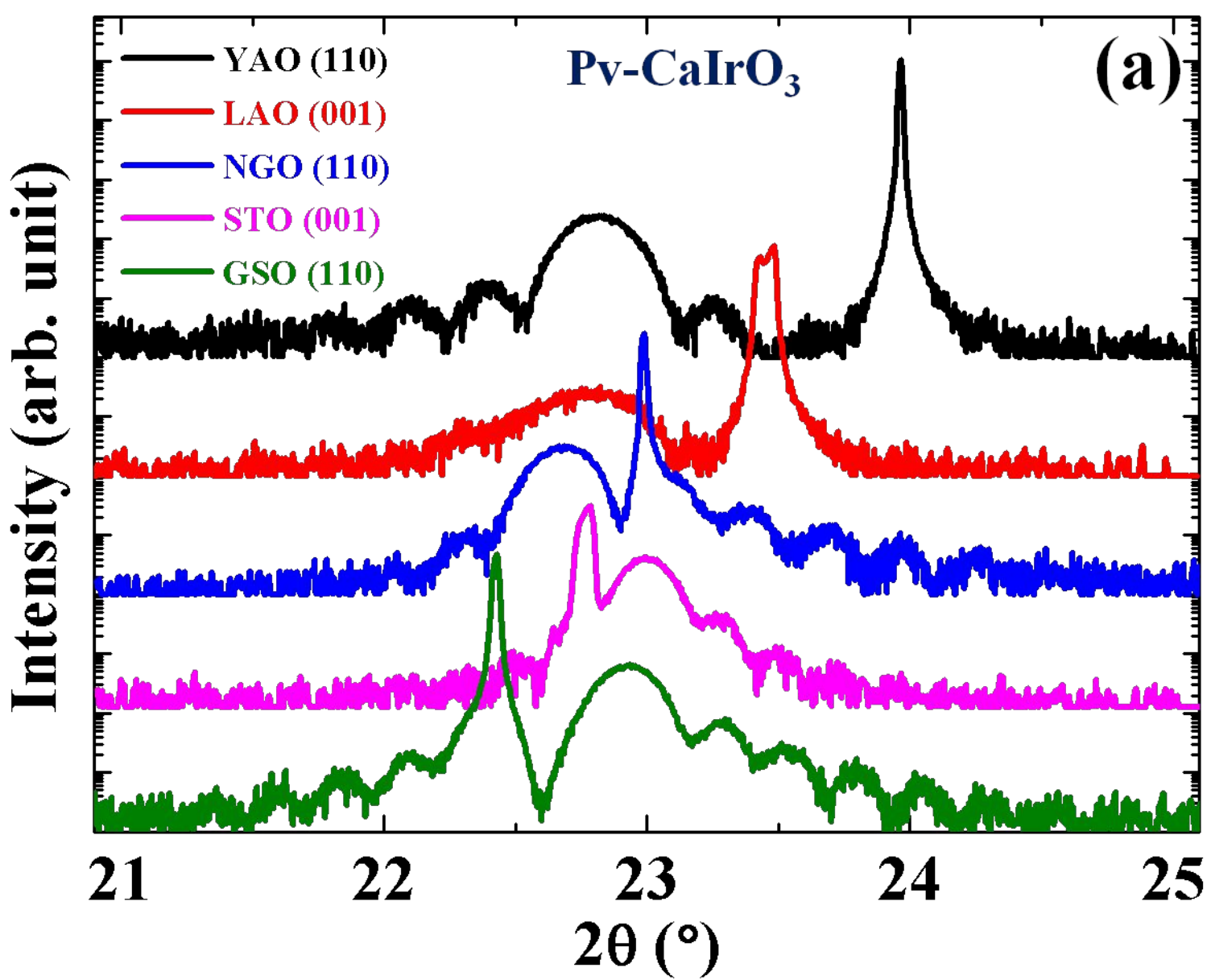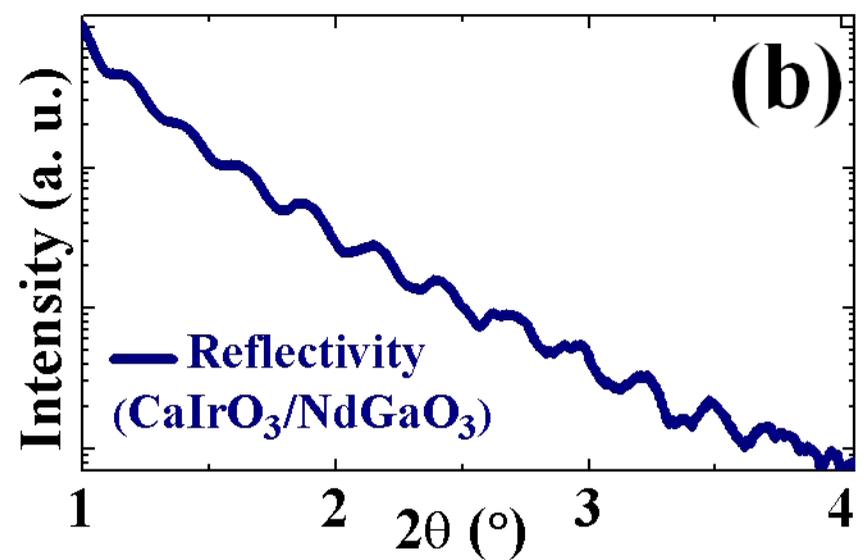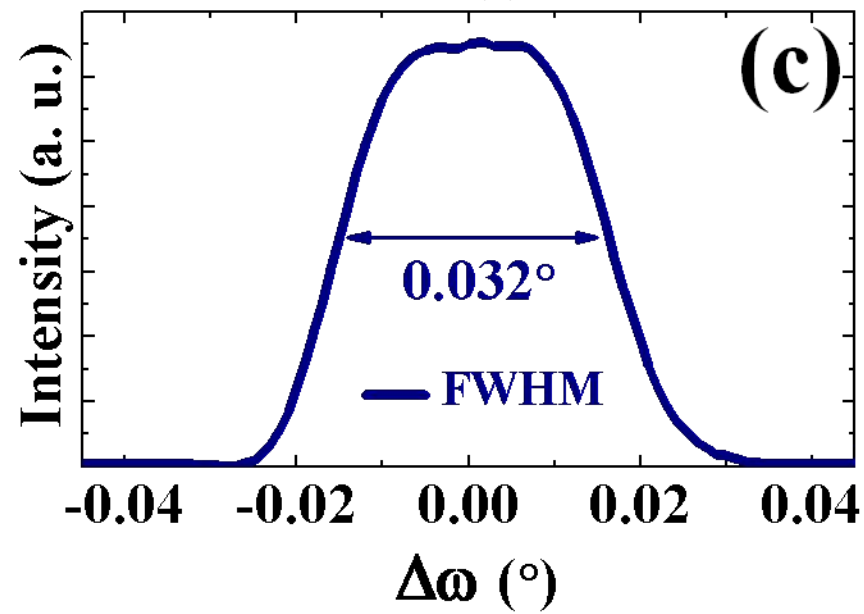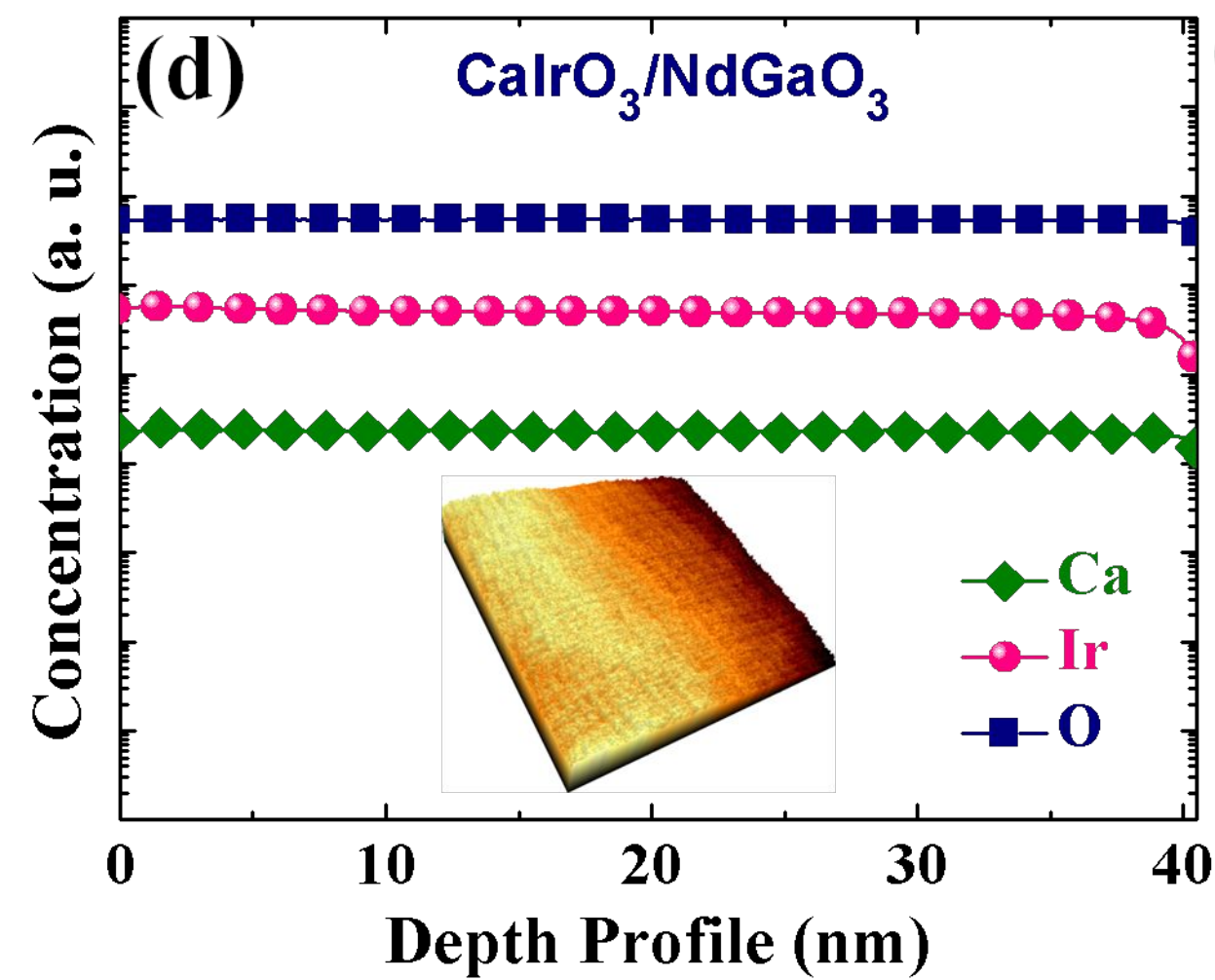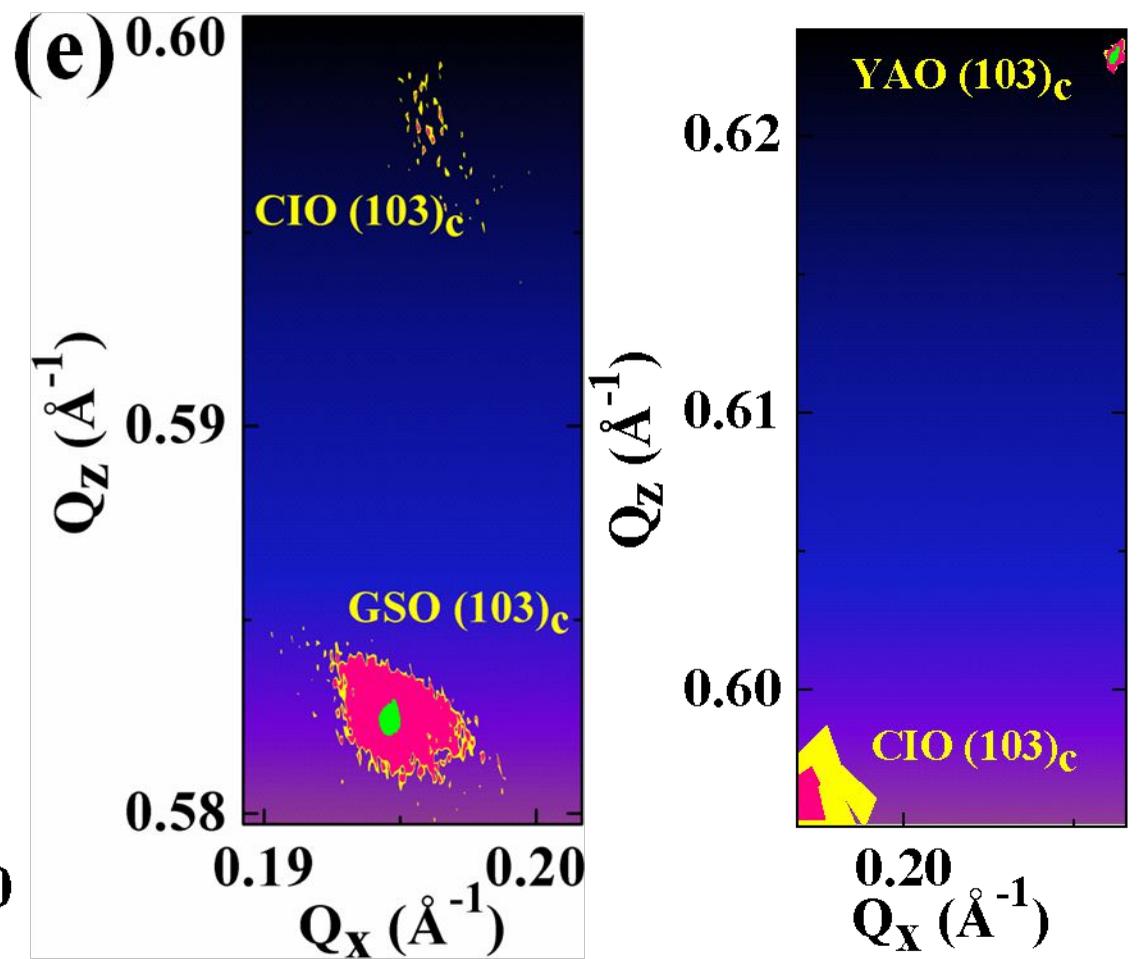

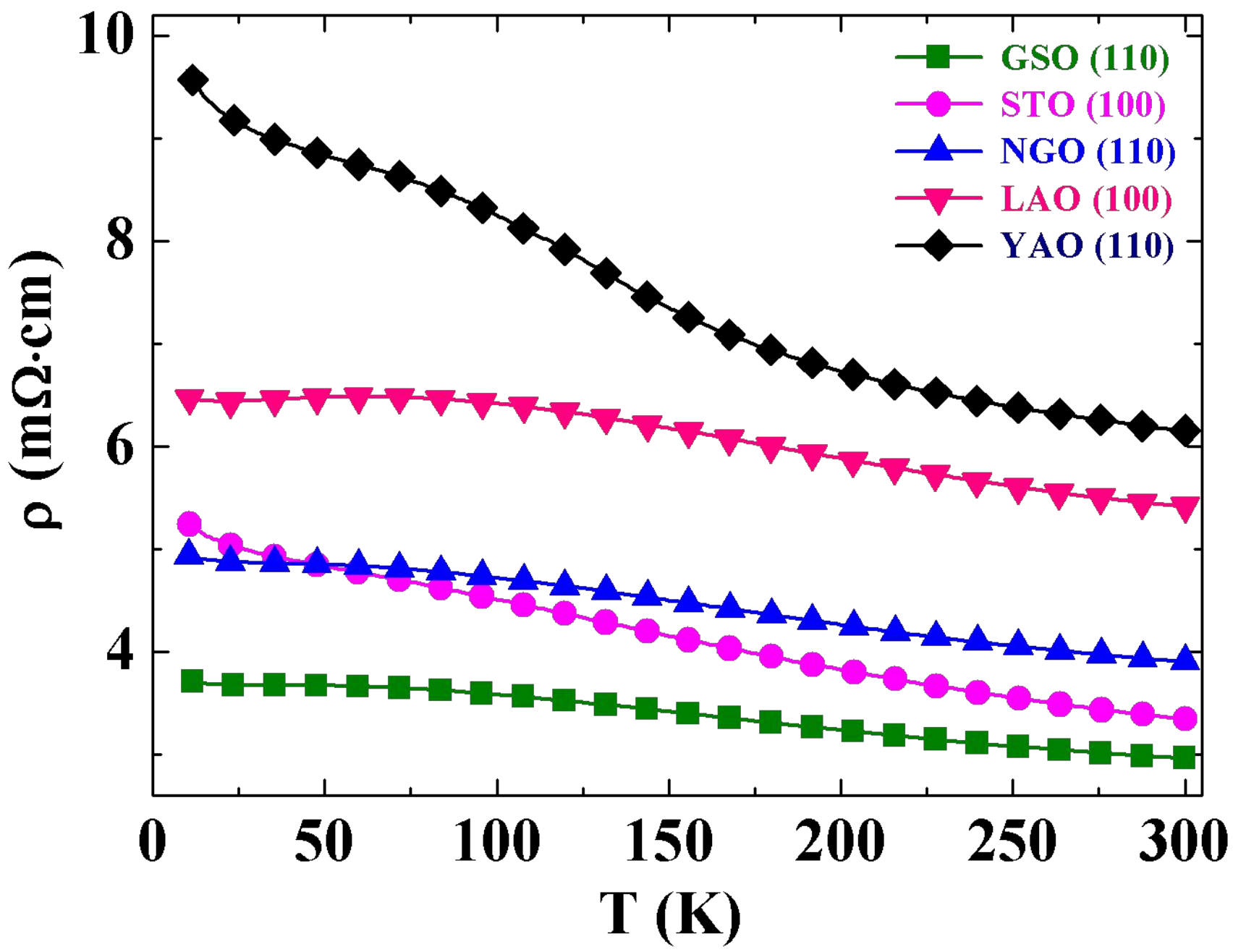

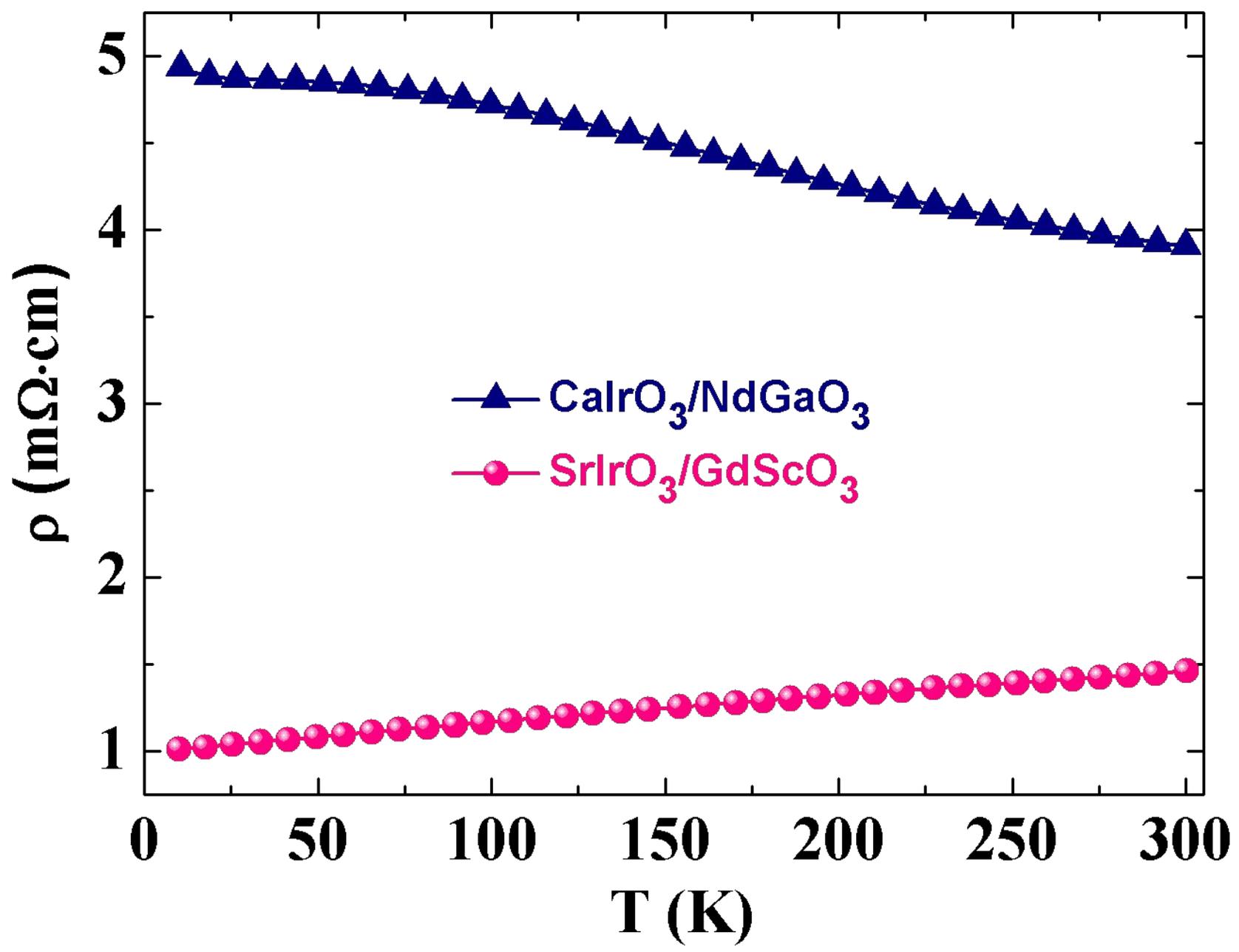

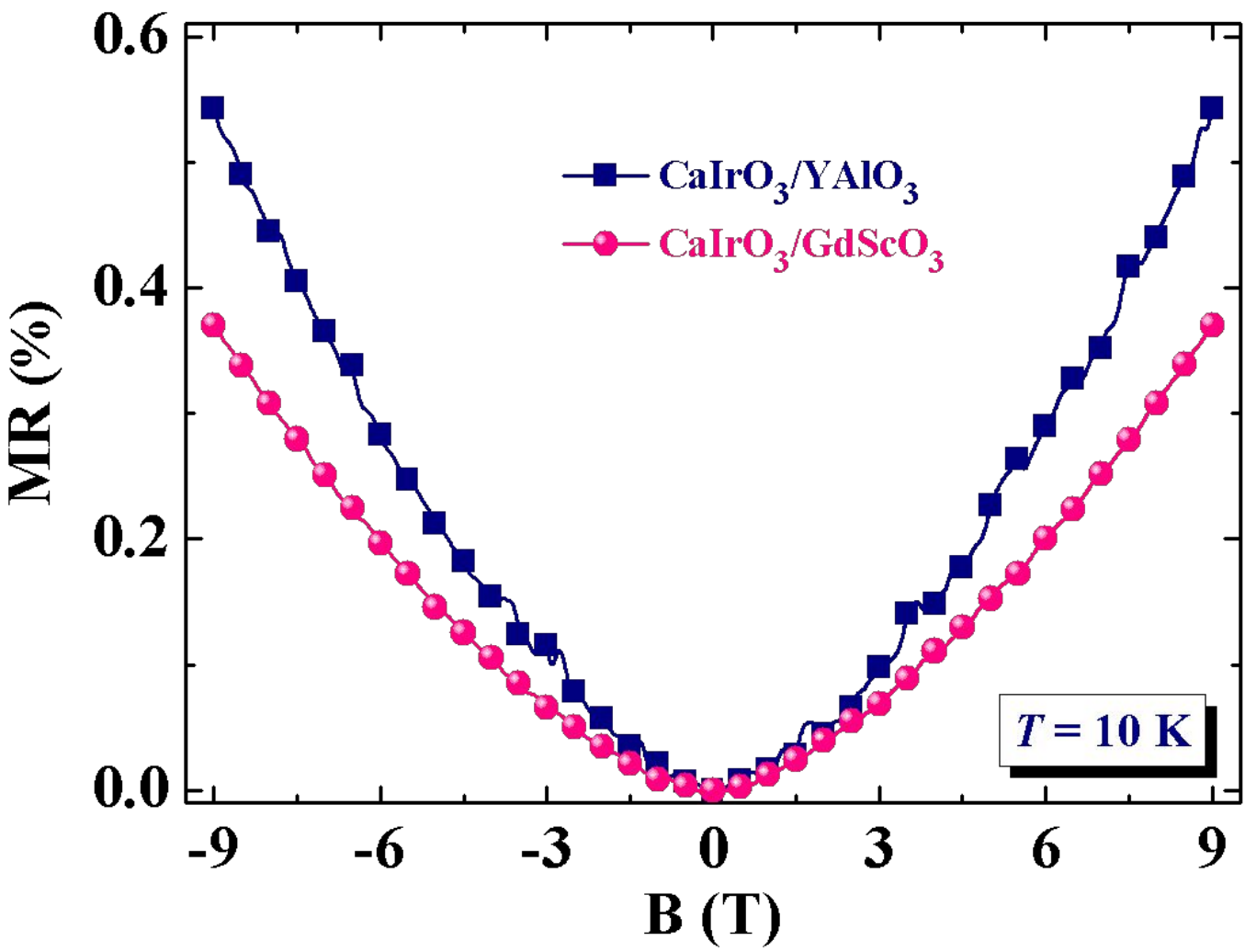